\documentclass[runningheads]{llncs}
\usepackage{graphicx}
\usepackage{booktabs}

\begin{document}
\title{Autopet Challenge 2023: nnUNet-based whole-body 3D PET-CT Tumour Segmentation}
\author{Anissa Alloula\inst{1}\orcidID{0000-0003-1525-3994}\and
Daniel R McGowan\inst{2,3}\orcidID{0000-0002-6880-5687} \and
Bart\l omiej W. Papie\.z\inst{1}\orcidID{0000-0002-8432-2511}}
\authorrunning{Alloula et al.}
\institute{Big Data Institute, University of Oxford, England \and
Department of Oncology, University of Oxford, England \and Department of Medical Physics and Clinical Engineering, Oxford University Hospitals NHS FT, Oxford, England.
\email{anissa.alloula@kellogg.ox.ac.uk}\\}
\maketitle              % typeset the header of the contribution
\begin{abstract}
Fluorodeoxyglucose Positron Emission Tomography (FDG-PET) combined with Computed Tomography (CT) scans are critical in oncology to the identification of solid tumours and the monitoring of their progression. However, precise and consistent lesion segmentation remains challenging, as manual segmentation is time-consuming and subject to intra- and inter-observer variability. Despite their promise, automated segmentation methods often struggle with false positive segmentation of regions of healthy metabolic activity, particularly when presented with such a complex range of tumours across the whole body. In this paper, we explore the application of the nnUNet to tumour segmentation of whole-body PET-CT scans and conduct different experiments on optimal training and post-processing strategies. Our best model obtains a Dice score of 69\% and a false negative and false positive volume of 6.27 and 5.78 mL respectively, on our internal test set. This model is submitted as part of the autoPET 2023 challenge. Our code is available at: https://github.com/anissa218/autopet\_nnunet

\end{abstract}
\section{Introduction}

Positron Emission Tomography / Computed Tomography (PET/CT) imaging is a vital element of the diagnostic process and clinical management for a wide range of solid tumours \cite{petct-imaging-cancer,fundamentals-pet}. PET imaging, usually conducted with Fluorodeoxyglucose (FDG) radiotracer, a glucose analogue, provides metabolic information of areas with high glucose consumption, and these can be indicative of the presence of tumours \cite{fundamentals-pet,pet-fdg-paper}. On the other hand, CT images provide anatomic information, which can help a clinician determine whether abnormal glucose uptake corresponds to a normal healthy metabolic organ or a malignant tumour \cite{petct-imaging-cancer}. The accurate detection and quantification of tumours is crucial for diagnosis, tumour staging and characterisation, and monitoring of disease evolution. Manual tumour segmentation is a costly and long process, which may be subject to intra- and inter- clinician variability \cite{ai-in-petct}. For these reasons, automated segmentation of these tumours is of particular importance. 

Previous research has shown that it is possible to accurately segment tumours with a variety of deep learning algorithms such as convolutional neural networks like the U-Net or transformer-based architectures \cite{unet-paper,segmentation-survey,ai-lymphoma,ai-promise-petct}. However, whole-body PET/CT tumour segmentation is a particularly challenging problem given the inherent multi-modality of the data as well as the extent of anatomical coverage and morphological variability of the tumours that can be present across the body\cite{challenge-report}. Notably, last year's challenge results showed accurate segmentation of tumours in whole-body PET/CT scans, with a Dice score of 0.79 for the winning model \cite{challenge-report}. However, performance of the algorithms on images acquired in a different hospital to the ones they were trained on was substantially inferior, particularly in terms of Dice accuracy \cite{challenge-report}. This year's challenge, Autopet-ii, aims to extend this work by focusing on generalisation of the model to other acquisition protocols and sites. Generalisation beyond a single scanner or acquisition site is challenging because of domain shift, for instance due to different image resolutions, varying levels of noise, and spatial variations \cite{seg-prostate,generalisation-deep-stacked-networks}. This is also hindered by the lack of publicly available segmented PET/CT datasets on which robust models can be trained. To this end, the Autopet-ii organisers provide an extensive dataset of whole-body PET/CT scans of patients with and without tumours \cite{dataset-ref}.

In this paper, we extend nnUNet, a semantic segmentation method which automatically adapts to a given dataset, and which has shown state-of-the art results across a wide range of medical image segmentation tasks \cite{nnunet_paper}. We also investigate different post-processing methods to the improve final prediction. Indeed, nnU-Net based methods consistently outperformed other architectures in last year's challenge \cite{challenge-report}.

\section{Methods}

\subsection{Data}

The challenge provided a set of 1016 whole-body PET/CT scans publicly available on The Cancer Imaging Archive (TCIA) as well as the mask as segmented by two radiologists \cite{dataset-ref}. These were acquired from University Hospital Tübingen and University Hospital of the LMU with Siemens Biograph mCT, mCT Flow and Biograph 64, GE Discovery 690 PET/CT scanners. 900 patients were involved; approximately half had no cancer, and the other half presented with histologically-proven malignant melanoma, lymphoma, or lung cancer. 

In addition, a hidden test-set of 205 images were used to evaluate the models, which was drawn in part from the same source distribution (1/4) and in part (3/4) from a different distribution. 5 of these images were used for preliminary testing, and the remainder will be used for final evaluation at the end of the challenge (not available to the authors at the time of submission).

One of the submitted models was also trained on a dataset including both the TCIA PET/CT scans and an additional 200 PET/CT head and neck scans from the Head and Neck Tumour Segmentation and Outcome Prediction in PET/CT 2022 challenge (HECKTOR) \cite{hecktor-challenge}. The aim was increase the diversity of the training data in order to improve model generalisability. 

\subsection{Pre-processing}
DICOM files were resampled (CT to PET imaging resolution, ranging from 200-677x400x400) and normalised. This involves Z-score intensity normalisation for PET images and global dataset percentile clipping and Z-score normalisation for CT. Subsequent pre-processing was done according to standard nnUNet procedure \cite{nnunet_paper}. The PET and CT images were concatenated into 2-channel 3D images with a median size of 2x236x400x400, and 90\% were kept for training and cross-validation, 10\% were kept as a held-out internal test set.

\subsection{Architecture and training}
A standard 3D full resolution U-Net was trained and cross-validated on a Tesla P100-SXM2-16GB GPU for 5 folds with the specifications shown in Table 1. PET-CT channels were concatenated and cropped to patches as input. For each batch of 2 patches, oversampling was implemented so that at least one of the two contained a positive label in the ground truth segmentation. This was done to ensure the model was trained with enough examples of lesions. 

\begin{table}[]
\centering
\caption{Final training strategy}\label{tab1}
\begin{tabular}{ll}
\toprule
\textbf{Training strategy}    & \textbf{Final Implementation}         \\ \midrule
Number of epochs              & 1500                                  \\ \hline
Iterations per epoch          & 250                                   \\ \hline
Batch size                    & 2                                     \\ \hline
Patch size                    & 128x128x128                           \\ \hline
Foreground batch oversampling & $>$50\%                                  \\ \hline
Learning rate                 & 0.0001             \\ \hline
Learning scheduler                 & Poly learning rate schedule                                         \\ \hline
Loss function                 & Soft dice loss and cross entropy loss \\ \bottomrule
\end{tabular}
\end{table}

Standard nnUNet data augmentation was used during training. This includes random rotation, scaling, Gaussian noise and blur, brightness, contrast augmentation, gamma correction, simulation of low resolution, and mirroring.

\subsection{Post-processing}

After training, different methods of post-processing were evaluated, based on the removal of predicted tumour areas of low size. This was done through connected component analysis of positive voxels \cite{cc3d}.

\subsection{Evaluation}
Inference was conducted on the held-out set of 101 images and all of the models were evaluated based on three metrics. At a later stage, the preliminary test set (released by the organisers) of 5 images was also used for evaluation. For the images where tumours were present, Dice overlap score of the segmented lesions, as well as volume of false positive connected components that do not overlap with true positives, were measured. In all images, false negative volume was also evaluated. This corresponds to the volume of positive connected components in the ground truth label that did not overlap with any positive segmented mask.

\section{Results}

% Initial data exploration

Models were trained with a variety of nnUNet parameters in order to determine the optimal configuration, and results are presented in Table \ref{tab2}. These results suggest that increasing training duration improved the model's performance on both the cross-validation and testing set. Removing mirroring as a data augmentation step also showed similar benefits, perhaps as there are important specifics to the organs on the left and right side of the PET/CT scans. Moreover, increasing the input patch size and maximum number of feature maps used in the model's architecture did not translate to better performance in the test set, perhaps due to over-fitting of the model. Finally, reducing the number of positive images presented to the network at each batch did not improve performance.

% separate test and CV in table
\begin{table}[]
\centering
\caption{Dice, false negative (FN), and false positive (FP) scores on cross-validation (CV) images across all folds, as well as on 101 held-out internal test images. Regions with fewer than 10 connected voxels were removed from test predictions. Inference was not performed for the final model \textit{g} due to time constraints.}\label{tab2}
\resizebox{12cm}{!}{%
\begin{tabular}{lcccccc|ccc|ccc}
\toprule
\multicolumn{7}{c|}{\textbf{Model hyper-parameters}} & \multicolumn{3}{c|}{\textbf{Cross-Validation}} & \multicolumn{3}{c}{\textbf{Internal Testing}} \\ \midrule
\textbf{Model} & \textbf{Epochs} & \textbf{Patch Size} & \textbf{Features} & \textbf{LR} & \textbf{Oversampling (\%)} & \textbf{Mirroring} & \textbf{Dice}   & \textbf{FN}      & \textbf{FP}      & \textbf{Dice}   & \textbf{FN}     & \textbf{FP}      \\ \midrule
baseline       & 1000            & 128                  & 256               & 0.001                  & 50                 & Yes                 & 0.716  & 8.345   & 11.916  & 0.664  & 7.651  & 11.883  \\ \midrule
a              & 1500            & 128                  & 256               & 0.001                  & 50                 & Yes                 & 0.734  & \textbf{7.430} & 10.258  & 0.680  & 7.376  & 8.043   \\ \midrule
b              & 1500            & 128                  & 256               & 0.001                  & 50                 & No                  & 0.732  & 8.265   & 8.577   & \textbf{0.685}  & \textbf{6.270}  & 5.778   \\ \midrule
c              & 1500            & 128                  & 256               & 0.01                   & 50                 & Yes                 & 0.670  & 9.926   & 22.255  & 0.611  & 14.878 & 13.408  \\ \midrule
d              & 1500            & 192                  & 512               & 0.01                   & 50                 & Yes                 & \textbf{0.743}  & 13.960  & \textbf{6.746}  & 0.673  & 13.807 & 3.788   \\ \midrule
e              & 1000            & 192                  & 512               & 0.001                  & 50                 & Yes                 & 0.716  & 8.345   & 11.916  & 0.682  & 8.923  & \textbf{3.239}   \\ \midrule
f              & 1000            & 128                  & 256               & 0.001                  & 33                  & Yes                 & 0.714  & 7.930   & 13.200  & 0.660  & 8.190  & 12.412  \\ \midrule
g              & 1000            & 128                  & 256               & 0.001                  & 10                  & Yes                 & 0.722  & \textbf{7.227}  & 23.790  & /      & /      & /       \\ \bottomrule
\end{tabular}%
}
\end{table}

\begin{figure}[ht]
  \centering
  \includegraphics[width=0.75\textwidth]{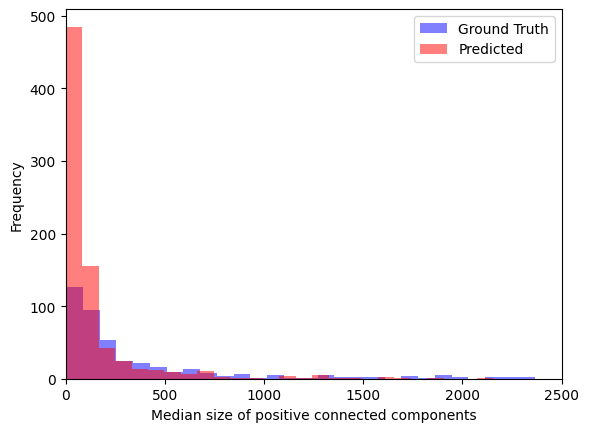}
  \caption{Distribution of median positive connected component size across all training images. The purple bars show the sizes of the components in the ground truth labels while the red bars show the median sizes predicted by model ``b''.}
  \label{fig:conn-comp-dist}
\end{figure}

The examination of the model's prediction showed that a high proportion of the segmentations were of small size (see Figure \ref{fig:conn-comp-dist}). As can be seen in Table \ref{tab3}, removing these tumours  caused a minor reduction in false positive volumes compared to no removal (min size of 0). A threshold of 10 appeared as optimal, maintaining a high Dice score while slightly lowering false positive scores.

\begin{table}[h!]
\centering
\caption{Effect of the removal of segmented lesions below a certain threshold size of connected components on the three challenge metrics in the test set.}\label{tab3}
\resizebox{5cm}{!}{%
\begin{tabular}{llll}
\toprule
\textbf{Min size threshold} & \textbf{Dice}  & \textbf{FP}    & \textbf{FN}     \\ \midrule
0                  & 0.685 & 5.778 & 6.270  \\ \midrule
5                  & 0.685 & 5.763 & 6.510  \\ \midrule
10                 & 0.686 & 5.744 & 6.723  \\ \midrule
20                 & 0.686 & 5.679 & 7.909  \\ \midrule
40                 & 0.653 & 5.576 & 9.195  \\ \midrule
80                 & 0.621 & 5.312 & 11.623 \\ \bottomrule
\end{tabular}%
}
\end{table}
\newpage
\section{Discussion}
As model ``b'' obtained the best results on the internal test set, it was submitted as our final model to the Autopet-2023 challenge. 

However, in the future, many aspects remain to be investigated. Firstly, the model outputs high volumes of false positive and false negative components, despite having a relatively high Dice overlap score. Ensembling the predictions of multiple models of different architectures and trained on different modalities of input data (for instance just PET, just CT, or both) may help maximise robustness and minimise false predictions. Indeed, in last year's challenge and other similar medical image segmentation tasks, ensembles of multiple diverse models often outperform any individual model \cite{challenge-report,diversity-ensembles,divergentnets}.

This includes developing an end-to-end model which predicts fewer very small false positive connected components and therefore does not require post-processing. Modifying the loss function, for instance with generalised Dice overlap, which weighs the contribution of each label by the inverse of its volume, may be a way to target this by penalising false small predictions \cite{generalised-dice-overlap}. 

Moreover, in order to build a more generalisable model, it would be of interest to quantify the uncertainty associated with the model's predictions. This would give an indication of how trustworthy each prediction is, which would be particularly useful when doing inference on out-of-distribution images \cite{brats-uncertainty,evidencecap-uncertainty,evidential-lymphoma}. Incorporating uncertainty during model training could also be beneficial by maximising the learning of examples which the model is uncertain about \cite{evidencecap-uncertainty}.

This work highlights the strength and adaptability of the nnUNet, which, with very little parameter tuning, accurately segmented lesions. This work and the Autopet-ii challenge represent crucial steps towards the development of reliable and robust PET/CT segmentation algorithms, with  significant potential for valuable clinical application.

\section{Acknowledgements} % do i want this as a numbered subsection

This work was supported by the EPSRC grant number EP/S024093/1 and the Centre for Doctoral Training in Sustainable Approaches to Biomedical Science: Responsible and Reproducible Research (SABS: R3) Doctoral Training Centre, University of Oxford. 
The authors acknowledge the AUTOPET challenge for the free publicly available PET/CT images used in this study.
The computational aspects of this research were supported by the Wellcome
Trust Core Award Grant Number 203141/Z/16/Z and the NIHR Oxford BRC. The views
expressed are those of the author(s) and not necessarily those of the NHS, the
NIHR or the Department of Health.

\newpage
\bibliographystyle{unsrt}

\bibliography{mybibliography}

\end{document}